# Electric Field Induced Cubic to Monoclinic Phase Transition in Multiferroic $0.65Bi(Ni_{1/2}Ti_{1/2})O_3$-$0.35PbTiO_3$ Solid Solution

*Rishikesh Pandey and Akhilesh Kumar Singh**

**Email: akhilesh_bhu@yahoo.com*

School of Materials Science and Technology, Indian Institute of Technology (Banaras Hindu University) Varanasi- 221005, India

The results of x-ray diffraction (XRD) studies on $0.65Bi(Ni_{1/2}Ti_{1/2})O_3$-$0.35PbTiO_3$ solid solution poled at various electric field is presented. After poling, significant value of planar electromechanical coupling coefficient ($k_P$) is observed for this composition having cubic structure in unpoled state. The cubic structure of $0.65Bi(Ni_{1/2}Ti_{1/2})O_3$-$0.35PbTiO_3$ transforms to monoclinic structure with space group Pm for the poling field $\geq$ 5kV/cm. Large c-axis microscopic lattice strain (1.6 %) is achieved at 30 kV/cm poling field. The variation of the c-axis strain and unit cell volume with poling field show a drastic jump similar to that observed for strain versus electric field curve in $(1-x)Pb(Mg_{1/3}Nb_{2/3})O_3$-$xPbTiO_3$ and $(1-x)Pb(Zn_{1/3}Nb_{2/3})O_3$-$xPbTiO_3$.

Recently, there has been great interest in multiferroic materials which exhibit both magnetic and ferroelectric ordering and show the presence of magnetoelectric coupling [1-3]. The magnetoelectric coupling provides additional degree of freedom in designing potential devices [1-2]. Existence of magnetism and ferroelectricity in materials usually has contradictory electronic configuration requirement and hence the magnetoelectric materials are rare [3]. Electrons in d-orbit of the transition metal favour the magnetism but reduce its tendency of hybridization with oxygen ions which is crucial for the appearance of ferroelectricity [4]. It is the reason, why such compounds are rare [3] and many of the known multiferroic materials exhibit this behaviour well below the room temperature [5]. The (1-x)$Bi(Ni_{1/2}Ti_{1/2})O_3$-x$PbTiO_3$ (BNT-PT) is a recently discovered multiferroic solid solution with attractive properties and exhibits the ferroelectric and ferromagnetic ordering at ambient temperature [6]. This raises the possibility of designing devices based on magnetoelectric coupling, operating at room temperature [5]. In the present work, we have discovered an electric field induced crystallographic phase transition in the pseudocubic composition of multiferroic BNT-PT solid solution which may be very promising in future for developing potential devices based on multiferroics. This phase transition is also accompanied by significant change in the unit cell volume and large microscopic strain (~1.6 %) in c-direction. This suggests that it is not necessary to use morphotropic phase boundary (MPB) compositions for getting large strain for micro-positioner and actuator applications. The pseudocubic compositions in the vicinity of the MPB can also give giant electromechanical response resulting from electric field induced transition. The present discovery calls for a systematic study of the electric field induced phase transitions and its correlation to the electromechanical responses in the various MPB solid solutions.

In recent years, electric field induced phase transitions have been investigated in several Bi-based ferroelectric solid solutions [7-12]. However, systematic investigation of electric field induced structural changes and its correlation to electromechanical response is still lacking for most of the MPB solid solutions. Electric field induced phase transition has also been investigated in Pb-based piezoelectric systems such as $PbZr_{1-x}Ti_xO_3$ (PZT) [13-14], $Pb(Mg_{1/3}Nb_{2/3})O_3$ [15], (1-x)$Pb(Mg_{1/3}Nb_{2/3})O_3$-x$PbTiO_3$ [16-17] and (1-x)$Pb(Zn_{1/3}Nb_{2/3})O_3$-x$PbTiO_3$ [18-19] etc. Change in the compositional width of the MPB is observed in (1-x)$Pb(Mg_{1/3}Nb_{2/3})O_3$-x$PbTiO_3$ [20] and (1-x)$Bi_{0.5}Na_{0.5}TiO_3$-x$BaTiO_3$ [21] after the application of electric field. The rhombohedral $0.8Na_{0.5}Bi_{0.5}TiO_3$-$0.2K_{0.5}Bi_{0.5}TiO_3$ is reported to transform into mixture of rhombohedral and tetragonal phases at applied electric field of more than 20 kV/cm [7]. In the lead free (1-x)$(Na,K)(Nb,Sb)O_3$-x$LiTaO_3$ solid solution, the structure and physical properties are reported to change on the application of electric field [22]. Rao et al. have shown that electric-field induced transformation from monoclinic (Cc) to rhombohedral (R3c) structure is observed in lead free $Na_{1/2}Bi_{1/2}TiO_3$ after poling [12]. Recently, domain wall reorientation has been investigated under applied electric field in 0.55BNT-0.45PT [23]. Electric field induced phase transitions from rhombohedral (R3c) to tetragonal phase is reported in $(Na_{0.5}Bi_{0.5})_{1-x}Ba_xTiO_3$ [24]. Ma et al. [25] have shown that the compositional width of the MPB can be modified by applying the electric field in (1-x)$Bi_{0.5}Na_{0.5}TiO_3$-x$BaTiO_3$ solid solution. Thus, the electric field induced phase transition appears to be a common feature of the MPB solid solutions for the compositions close to MPB where the phase stability is very sensitive to small stimulus like electric field, pressure or compositional change. In the present investigation, we have analysed the structure of 0.65BNT-0.35PT solid solution poled at different electric field strength to study the electric field induced phase transition.

Samples were synthesized by traditional solid state ceramic method. Stoichiometric mixture of analytical reagent grade powders $Bi_2O_3$, PbO, NiO and $TiO_2$ were used as raw material. The details of sample preparation are reported elsewhere [26]. For the study of electric field induced phase transition, the pellets were electroded by coating fired on silver paste cured at 500 $^0$C for 5 min. The electroded pellets were poled in silicone oil bath at 100 $^0$C for 15 min by applying DC field and then field cooled from 100 $^0$C to room temperature. Planar electromechanical coupling coefficient ($k_P$) was calculated by resonance-antiresonance method [27]. X-ray diffraction (XRD) measurement was carried out using an 18 kW rotating Cu-anode based Rigaku (Japan) powder diffractometer operating in the Bragg-Brentano geometry and fitted with a curved crystal graphite monochromator. To record the XRD data from poled samples, silver electrode was removed from one surface of the pellet. Structure was studied by Rietveld and Le-Bail full pattern matching analysis [28]. The anisotropic peak shape function suggested by Stephens and also incorporated in the Fullprof program was used to model the anisotropic XRD profile broadening during refinement [28]. The March function incorporated in Fullprof program was used to model the preferred orientation of the XRD profiles [28]. The refined value of the preferred orientation parameter [28] was close to 1 suggesting insignificant preferred orientation in the XRD patterns of the poled samples.

Fig.1 shows the powder XRD profiles of BNT-PT solid solution for the compositions with x=0.35, 0.41, 0.43, 0.49 and 0.55 in the 2θ range of $20^0$ to $60^0$. The indices shown on the peaks in Fig.1 correspond to tetragonal cell. For the compositions with x=0.55, the pseudocubic reflection (200) is seen to be doublet while the (111) profile is a singlet. This characterizes the tetragonal structure for the composition with x=0.55, as reported by us, very recently [26]. For the compositions with x=0.41 and 0.43, there is no prominent splitting but clear asymmetric shoulder towards the lower 2θ side is seen for the (200) profile. We have reported recently that the structure of these two compositions is monoclinic in the Pm space group. Detailed structural analysis of the MPB phase in BNT-PT solid solution using Rietveld method can be found in Ref.26. For the composition with x= 0.35, within the resolution of the laboratory XRD data, all the profiles are singlet and reveal the cubic structure with space group $Pm\bar{3}m$. The cubic structure of the composition with x=0.35 is confirmed by Rietveld analysis of the powder XRD data collected at room temperature. Fig.2 depicts the observed, calculated and difference profiles for the composition with x= 0.35 using space group $Pm\bar{3}m$ obtained by the Rietveld analysis of the XRD data. The fit between the observed and calculated profiles is quite good confirming the cubic structure. Inset to Fig.2 illustrates the goodness of fit for the (200) profile. The refined cubic lattice parameter 'a', $\chi^2$ and $R_{wp}$ (%) comes out to be 3.9657(2) Å, 2.54 and 14.0, respectively. The refined isotropic thermal parameters for A-site, B-site cations and oxygen anions were 4.0(7) Å$^2$, 2.3(1) Å$^2$ and 3.0(4) Å$^2$, respectively, which are quite high suggesting the large disorder in the structure. In view of this, we will call the structure of 0.65BNT-0.35PT to be pseudocubic, in the following discussions. High resolution synchrotron XRD data will be needed to unambiguously confirm the cubic structure of 0.65BNT-0.35PT.

Fig.3 shows the variation of planar electromechanical coupling coefficient ($k_P$) with composition (x). As can be seen from this figure, a peak is observed around the composition with x=0.45 corresponding to MPB. Most surprisingly, significant value of $k_P$ is obtained for the pseudocubic composition (x=0.35) also, which is not expected. This indicates that a phase transformation from pseudocubic to non-centrosymmetric ferroelectric phase occurs after poling. To confirm the electric field induced phase transition during poling, we carried out a detailed structural analysis of 0.65BNT-0.35PT samples poled at various electric fields. Fig.4 shows the XRD patterns of pellets poled at electric field 5 kV/cm, 10 kV/cm, 15 kV/cm, 20 kV/cm and 30 kV/cm. The XRD profiles of the poled samples show clear splitting in contrast to the singlet character for unpoled sample shown in Fig.1. To illustrate it further a comparison of the pseudocubic (110), (111) and (200) XRD profiles of the unpoled (continuous red line) and poled samples (scattered dots) at the poling field strength of 20 kV/cm is presented in Fig.5. As can be seen from this figure, there is drastic modification in the nature of the XRD profiles after poling. This clearly confirms an electric field induced structural phase transition in 0.65BNT-0.35PT. Splitting observed in (110), (111) and (200) pseudocubic XRD profiles of poled samples suggests the appearance of a monoclinic structure or coexistence of the monoclinic and tetragonal structures after poling, reported recently for the MPB compositions [26]. To determine the true crystallographic symmetry of the structure for the poled samples, we carried out Le-Bail full pattern matching analysis of the XRD data using different plausible space groups, i.e. rhombohedral $R\bar{3}m$, monoclinic Pm and monoclinic Cm. For the space group $R\bar{3}m$, the mismatch between observed and calculated profiles are quite prominent for (200) and (220) pseudocubic reflections. Similarly, for the space group Cm, the mismatch between the observed and calculate profiles are quite large for (200) and (220) pseudocubic reflections. The most satisfactory fit between the observed and calculated profiles is obtained for the Pm space group. There is no need to consider a coexisting tetragonal phase. A similar profile modification and electric field induced phase transition is reported in $0.93Bi_{0.5}Na_{0.5}TiO_3$-$0.07BaTiO_3$ but the symmetry of the resulting phase is tetragonal [10]. Fig.6 depicts the observed, calculated and difference profiles in the entire 2θ range for 0.65BNT-0.35PT sample poled at 30 kV/cm. As can be seen from this figure, the fit is quite satisfactory.

The refined lattice parameters using Pm space group obtained for 0.65BNT-0.35PT pellets poled at various fields are plotted in Fig.7. The lattice parameters ‘c’, ‘b’ and monoclinic angle (β) show slightly increasing trend with increasing the field strength, which suggests that the monoclinic distortion of the unit cell is increasing at higher poling fields. Inset Figs.7 (a) and (b) depict the variation of c-axis strain and unit cell volume, respectively, with poling field (E). The c-axis strain abruptly increases up to 1.14 % under the applied poling field of 5 kV/cm which gradually increases upto 1.6 % at the poling field of 30 kV/cm. The large strain results from the electric field induced pseudocubic to monoclinic phase transition. A similar strain jump is reported in PZN-PT [19, 29] and PMN-PT [29] subjected to external electric field. Park et al have reported that in single crystal 0.92PZN-0.08PT ultrahigh strain (~ 1.7 %) could be achieved at the electric field strength of ~ 120 kV/cm [29]. In the present investigation the strain level of ~ 1.6 % is obtained at significantly lower field strength (~ 30 kV/cm). It should be noted that 1.6 % strain reported here is the microscopic lattice strain along c-axis. The strain is expected to be lower in polycrystalline ceramic samples. However, in single crystal sample, one may expect to get this level of strain along c-axis at microscopic as well as macroscopic level. As shown in inset Fig.7 (b), the unit cell volume (V) also shows a jump with increasing poling field in the beginning (~ 10 kV/cm) and then levels off at higher fields. Significant increase in the unit cell volume is obtained due to the growth of the ferroelectric order during pseudocubic to monoclinic phase transition. Increased unit cell volume due to electric field induced phase transition is observed in (1-x)($Bi_{0.5}Na_{0.5}$)$TiO_3$-x$BaTiO_3$ also [21]. In PZT, Guo et al [13] have shown that both the rhombohedral and tetragonal compositions close to MPB exhibit cell distortions corresponding to monoclinic structure after poling at higher electric field. The first principles calculations [30] for PZT have shown that the monoclinic phase is the stable ground state in the MPB region. Transformation of the pseudocubic phase in to monoclinic structure after applying the electric field in the present work suggests that the monoclinic structure is the ground state for the compositions close to the MPB in the BNT-PT solid solution, also. Daniels et al [10] have proposed the segregation of grains into ferroelectric domains in pseudocubic 0.93$Bi_{0.5}Na_{0.5}TiO_3$-0.07$BaTiO_3$ on application of electric field. The preferred orientation parameter ($G_1$) [28] obtained by full pattern matching analysis of the XRD data for 0.65BNT-0.35PT comes out to be 1.001(3), 1.003(5), 1.009(4), 1.046(4) and 1.049(4) at the poling field of 5 kV/cm, 10 kV/cm, 15 kV/cm, 20 kV/cm and 30 kV/cm, respectively. The value of $G_1$ should be equal to 1 for no preferred orientation whereas it is <1 for platy habit and >1 for needle-like habit for the XRD data from the x-ray powder diffractometer with Bragg-Brentano geometry [28]. This suggests that the preferred orientation is not significant in poled samples. In future, it will be interesting to investigate the growth of any rhombohedral structure before the appearance of the monoclinic phase by in-situ measurements on 0.65BNT-0.35PT. Removing the poling field does not revert back the original pseudocubic phase and crystal properties in our sample. The polarization-electric field (P-E) measurement on the poled sample shows well saturated hysteresis loop suggesting that the polarization can be reversed by external electric field. A similar type of irreversible phase transformation is observed in the lead free solid solution 0.93$Bi_{0.5}Na_{0.5}TiO_3$-0.07$BaTiO_3$ and 0.2$Bi_{0.5}K_{0.5}TiO_3$-0.8$Bi_{0.5}Na_{0.5}TiO_3$ [7, 10, 31], also. Meanwhile, on heating the poled samples of BNT-PT, the structure reverts back to original pseudocubic phase. Similar type of phase reversal to pseudocubic phase is observed on crushing the poled pellets. In contrast, Rao et al have reported that the $Bi_{0.5}Na_{0.5}TiO_3$ ceramic retain the electric field induced rhombohedral phase even after crushing the poled sample [12].

In conclusion, the pseudocubic ($Pm\bar{3}m$) phase of 0.65BNT-0.35PT transforms into ferroelectric monoclinic phase (Pm) after poling and shows the piezoelectric response. The unit cell volume drastically changes after poling due to growth of ferroelectric order with the appearance of the monoclinic structure. Large microscopic c-axis lattice strain (1.6%) resulting due to this phase transition can be exploited for various micro-positioners and actuator applications. Our present study reveals that in addition to MPB compositions, the compositions in the vicinity of the MPB should also be explored for piezoelectric applications as the structure and physical properties may significantly change after poling.

Authors are thankful to Prof. Dhananjai Pandey, School of Materials Science and Technology, IIT (BHU) Varanasi, India for extending laboratory facilities. R.P. acknowledges university grant commission (UGC), government of India for financial support as senior research fellow (SRF).

**Figure captions:**

**Fig.1** Powder XRD profiles of BNT-PT solid solution for the compositions with x=0.35, 0.41, 0.43, 0.49 and 0.55 in the 2-theta range of $20^0$ to $60^0$. The indices shown on the peaks correspond to tetragonal cell.

**Fig.2** Observed (dots), calculated (continuous line) and difference (continuous bottom line) profiles obtained after Rietveld analysis using cubic space group $Pm\bar{3}m$ for unpoled 0.65BNT-0.35PT. Inset illustrates the quality of fit for the (200) profile.

**Fig.3** Variation of planar electromechanical coupling coefficient ($k_P$) with composition for BNT-PT solid solution.

**Fig.4** Evolution of XRD profiles for 0.65BNT-0.35PT samples poled at different electric fields. The indices shown on the peaks correspond to pseudocubic cell.

**Fig.5** A Comparison of XRD profiles for unpoled (continuous red line) and poled (scattered dots) samples of 0.65BNT-0.35PT at 20kV/cm.

**Fig.6** Observed (dots), calculated (continuous line) and difference (continuous bottom line) profiles obtained after Le-Bail full pattern matching analysis of the powder XRD data using Pm space group for 0.65BNT-0.35PT poled at 30 kV/cm. The vertical tick-marks above the difference plot show the peak positions.

**Fig.7** Variation of monoclinic lattice parameters with poling field (E) for 0.65BNT-0.35PT. Insets show the variation of (a) c-axis microscopic lattice strain (%), and (b) unit cell volume (V), with poling field (E) for 0.65BNT-0.35PT.

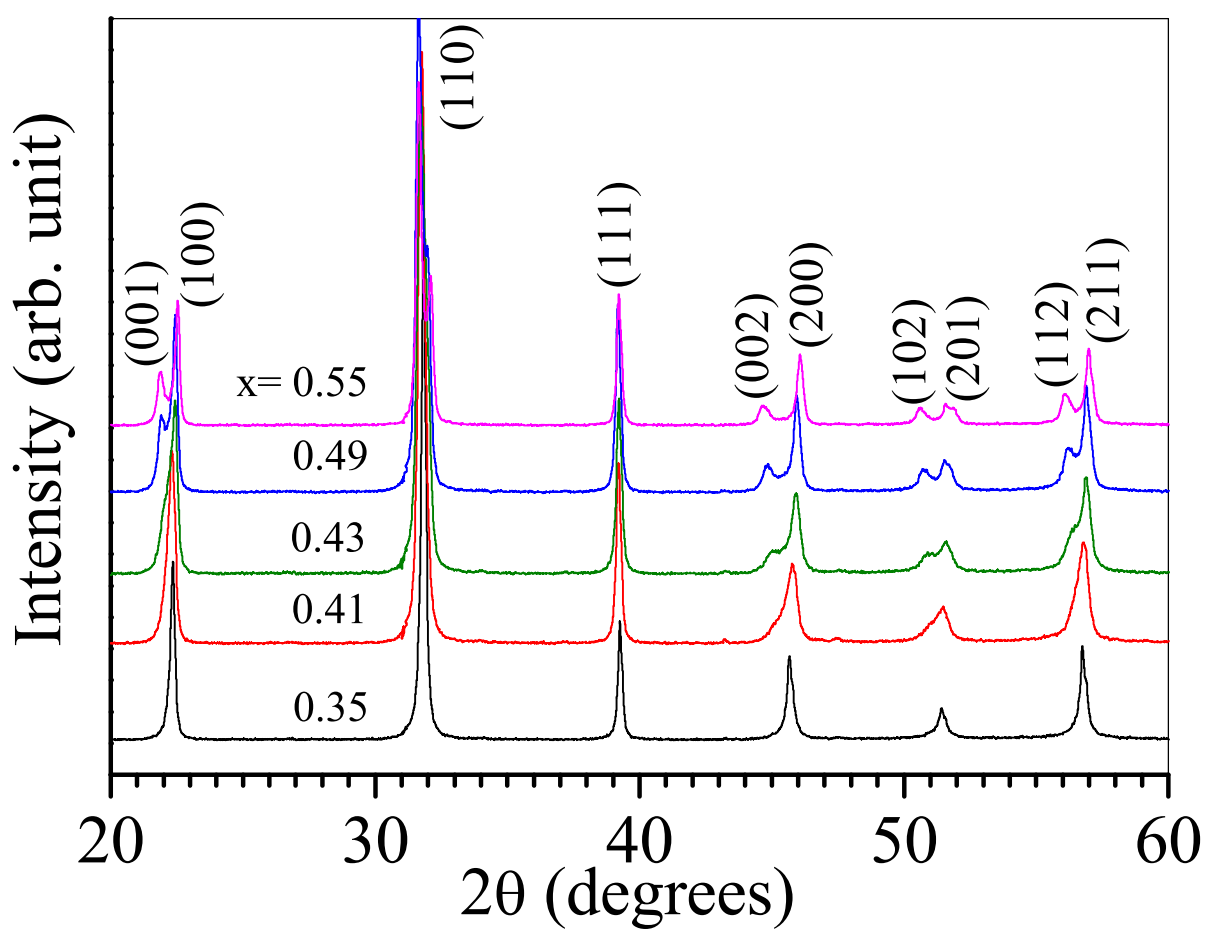

Fig.1

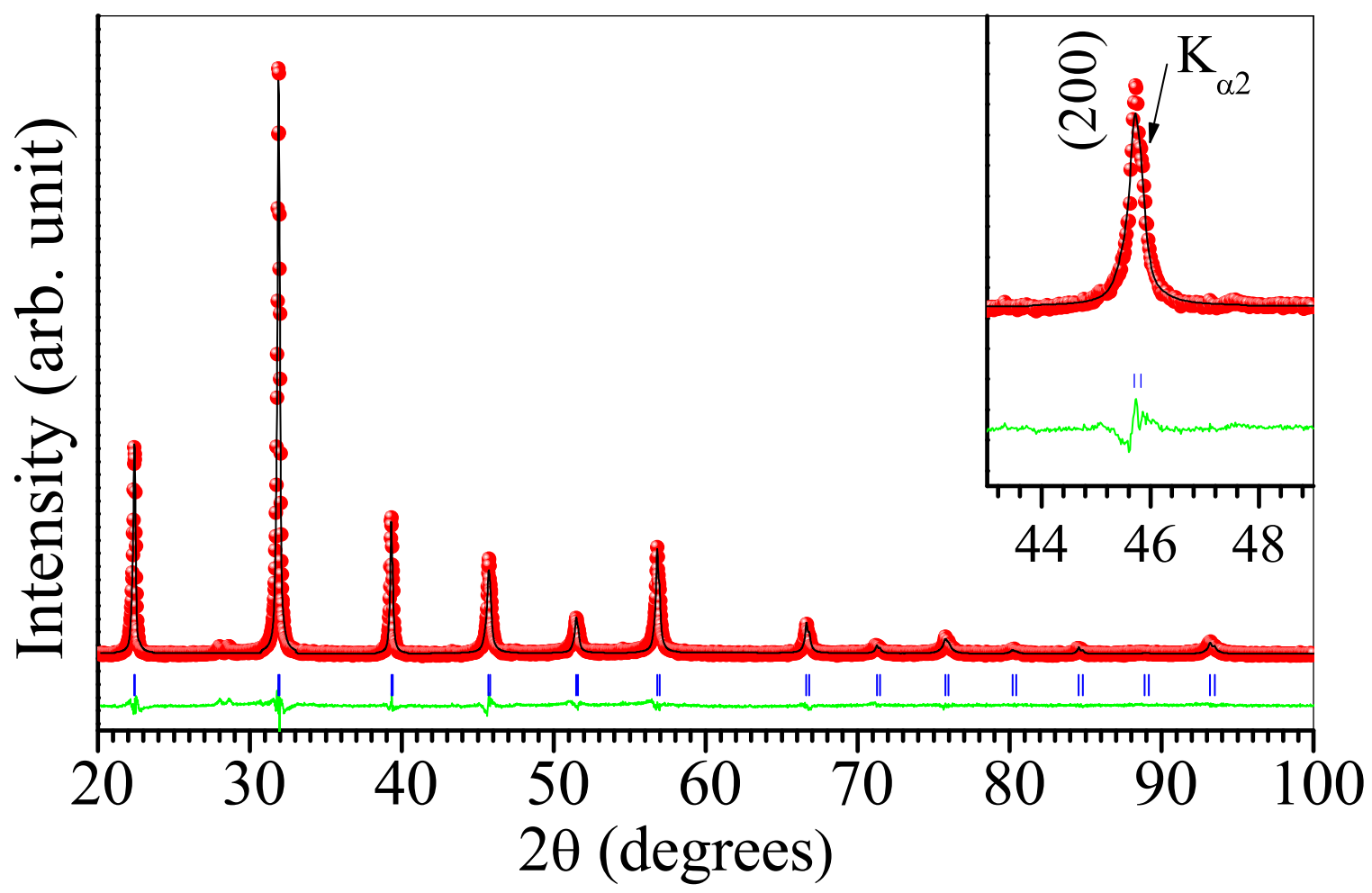

Fig.2

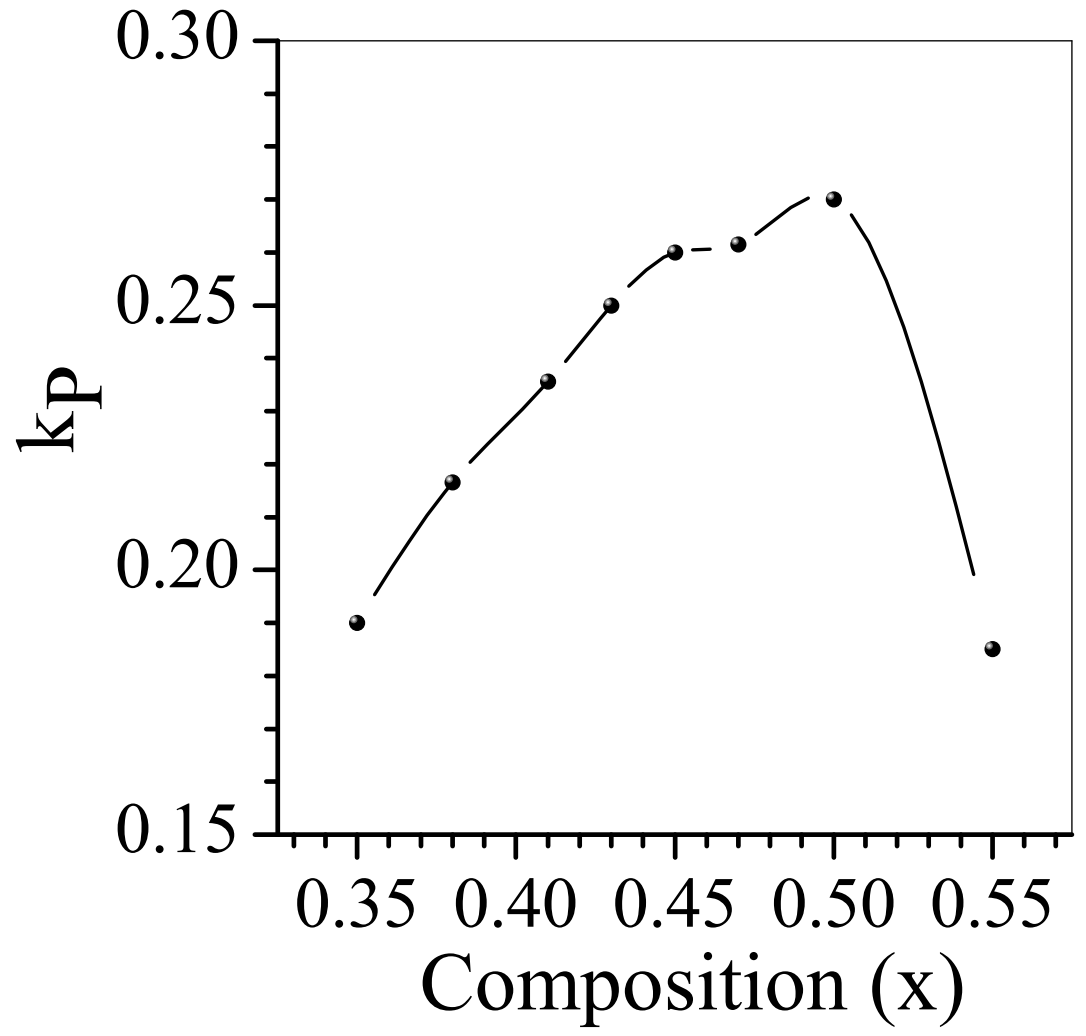

Fig.3

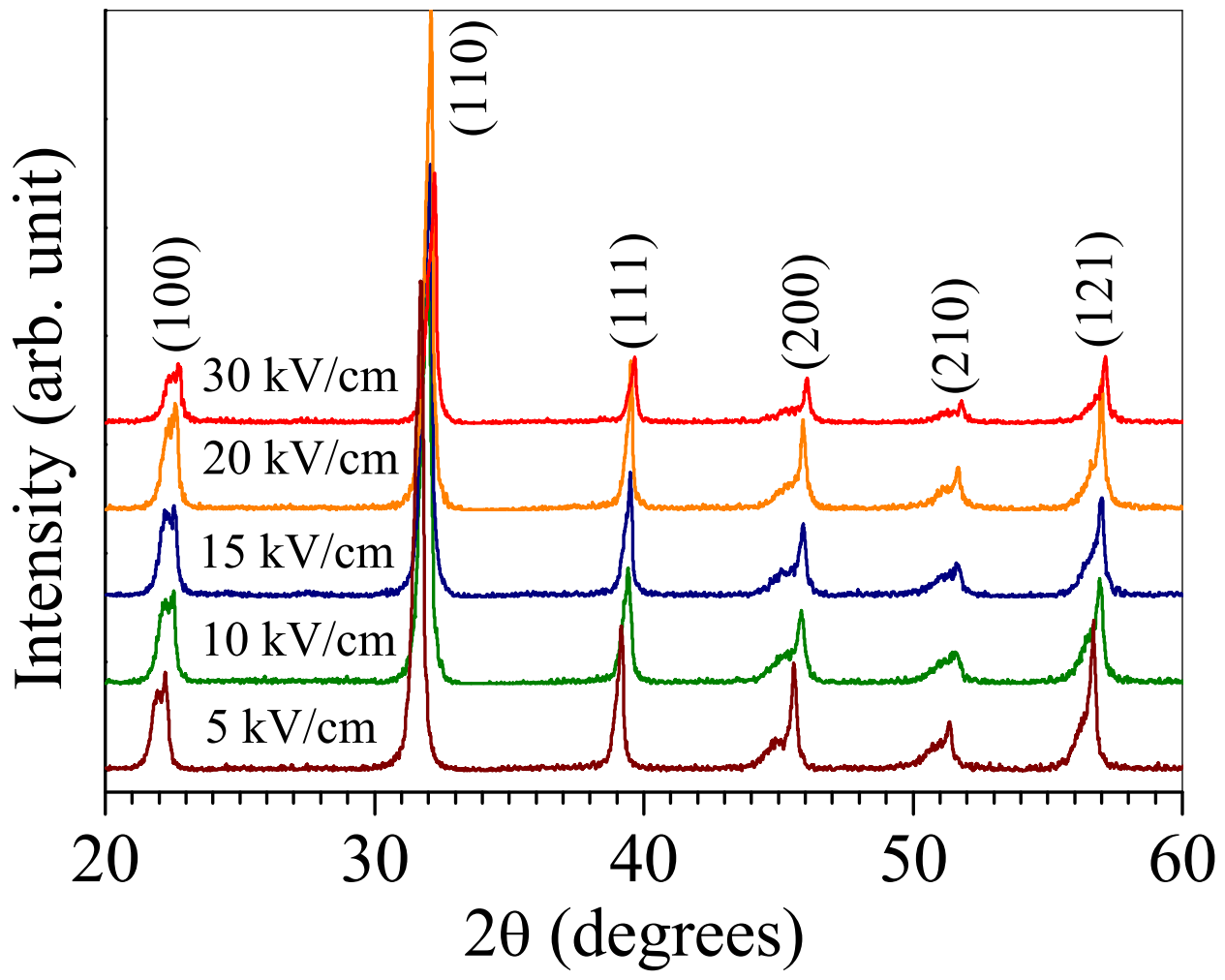

Fig.4

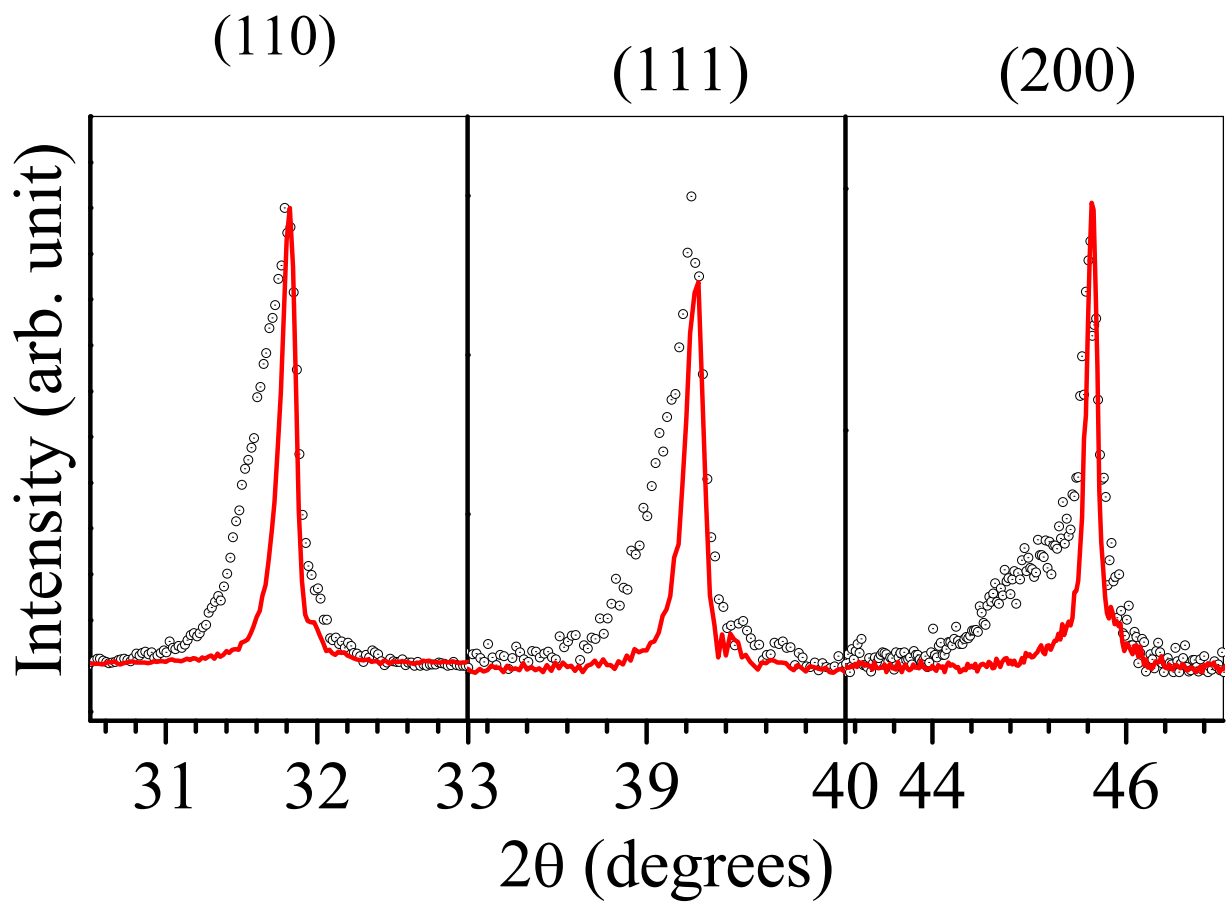

Fig.5

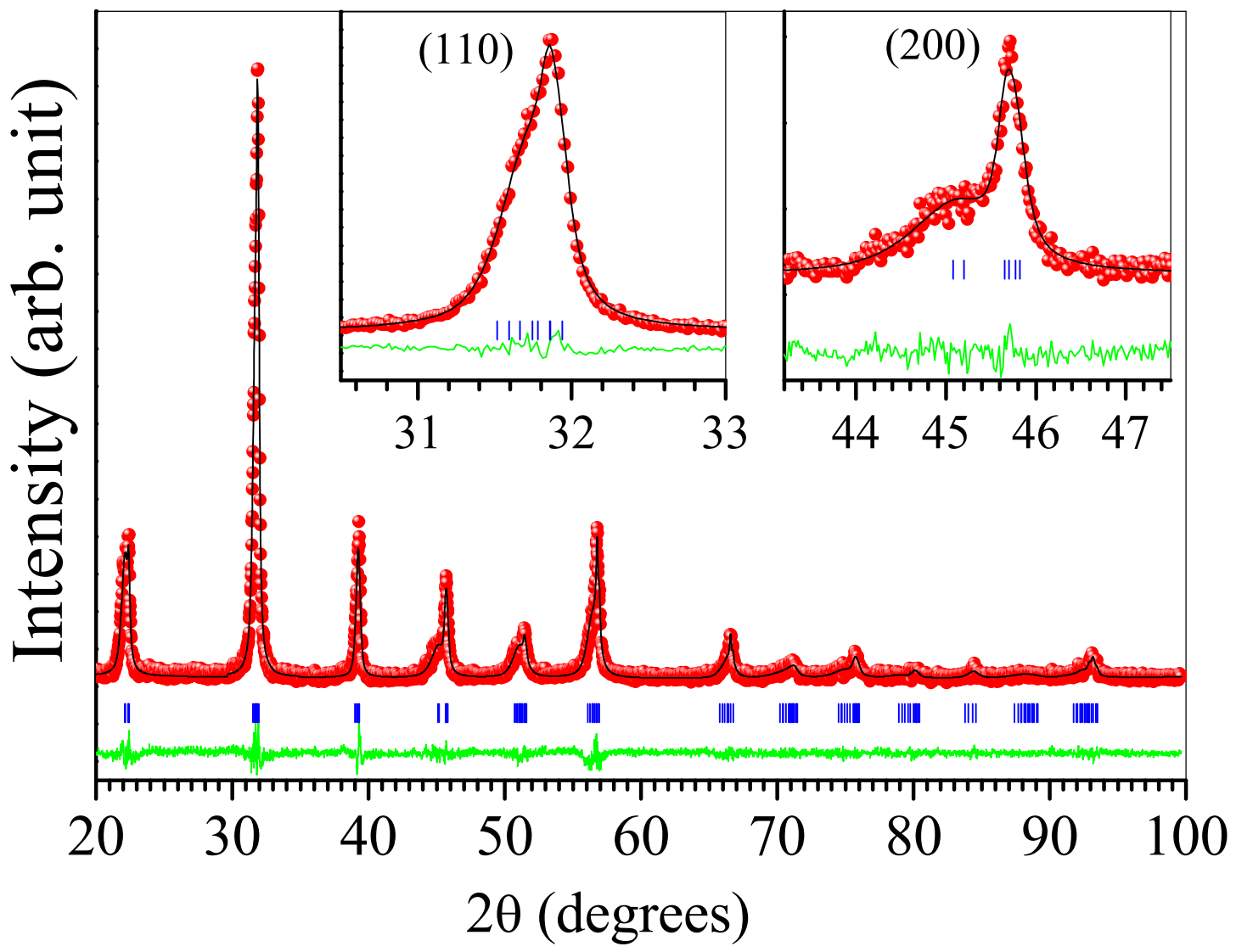

Fig.6

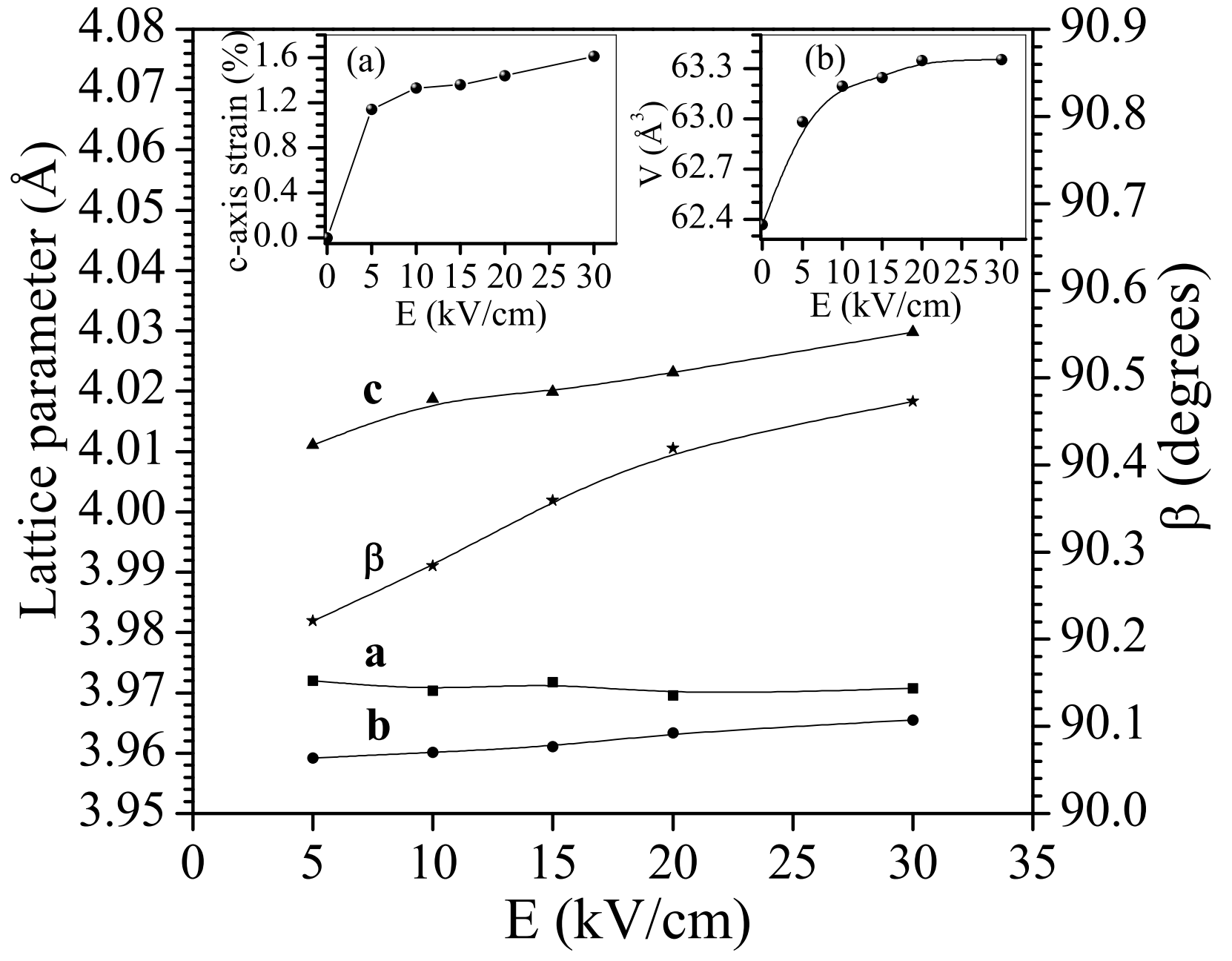

Fig.7